\documentclass[aps,prb,twocolumn,groupedaddress,amssymb,superscriptaddress]{revtex4}
\usepackage[dvips]{graphicx} % per latex (fig .ps)
\usepackage{bm}
\newcommand{\CLBLCO}{$\mathrm{Ca_xLa_{1-x}Ba_{1.75-x}La_{1.25+x}Cu_3O_{6+y} }$}

\begin{document}

\title{Experimental evidence of chemical-pressure-controlled
superconductivity in cuprates}

\author{S. Sanna}
\affiliation{Dipartimento di Fisica and CNISM, Universit\`{a} di Pavia, Via Bassi 6, 27100 Pavia }
\affiliation{Dipartimento di Fisica and CNISM, Universit\`{a} di Parma, viale Usberti 7A, 43100 Parma }
\author{S. Agrestini}
\affiliation{Laboratoire CRISMAT, UMR 6508, ISMRA, Boulevard du Marechal Juin, 14050 Caen, France}
\author{K. Zheng}
\affiliation{Dipartimento di Fisica and CNISM, Universit\`{a} di Parma, viale Usberti 7A, 43100 Parma }
\author{R. De Renzi}
\affiliation{Dipartimento di Fisica and CNISM, Universit\`{a} di Parma, viale Usberti 7A, 43100 Parma }
\author{N. L. Saini}
\affiliation{Dipartimento di Fisica, Universit\`{a} di Roma ``La Sapienza", P. le Aldo Moro 2, 00185 Roma, Italy}

\date{\today}

\begin{abstract}
X-ray absorption spectroscopy (XAS) and high resolution X-ray
diffraction are combined to study the interplay between electronic and
lattice structures in controlling the superconductivity in cuprates with a model
charge-compensated \CLBLCO\ (0$\leq x<$0.5, $y\approx$7.13) system.  In spite
of a large change in T$_{c}$, the doped holes, determined by the Cu L
and O K XAS, hardly show any variation with the $x$. On the other hand,
the CuO$_{2}$ plaquette size shows a systematic change due to
different size of substituted cations. The results provide a {\em
direct} evidence for the chemical pressure being a key parameter for
controlling the superconducting ground state of the cuprates.

\end{abstract}


\maketitle

%\section{Section title}
Even after two decades, high T$_{c}$ superconductivity (HTcS) in
cuprates continues to be an area of frontier research in condensed
matter physics due to the still unknown mechanism of this quantum
matter.  It is well established \cite{Tallon1995} that the transition
temperature, T$_{c}$, depends on the hole density $h$, showing a dome
shaped behaviour, with a maximum $T_c$ ($T_{c}^{max}$) being at an
optimum doping ($h^{opt}$ $\sim$0.16 holes per CuO$_2$ in each copper
plane). However, a debatable question is related to the cause of a
significant $T_{c}^{max}$ variation from one family to the other (from
few tens of Kelvin up to a maximum of 135 K) at constant hole density.
There is a growing number of experimental evidences
\cite{Muller1999,Bianconi2000,Agrestini2003,Bianconi2001,Fratini2008,
Slezak2008,Locquet1998,Abrecht2003,Gozar2008,Bozovic2002,Oyanagi2007,
Saini2001,Sato2000} that, beyond
the hole doping, the CuO$_2$ plane structure plays an important role,
depending on the lattice mismatch with the intercalated layers. This
is analogous to what has been known for the manganites
\cite{Hwang1995,Millis1998,Ahn2004,Dagotto2005}; however, the role of
the lattice in cuprates is still unclear since, on one hand it is
difficult to experimentally separate the lattice effect from that of
the hole doping within the same family compound, and on the other hand
it is almost impossible to compare different cuprate families, due to
the large difference of their structural properties. The
charge-compensated compound \CLBLCO (CLBLCO)
\cite{Goldschmidt1993}, offers a rare possibility to explore
simultaneously the doping dependence of $T_{c}$ as well as the origin
of the family dependence of $T_{c}^{max}$.  CLBLCO belongs to the
so-called 123 family, but the partial substitution of La on the Ba
site hampers the formation of extended Cu(1)O$_y$ chains (e.g. typical
of $\mathrm{YBa_2Cu_3O_y}$) stabilising the system in a tetragonal
phase.  In addition, unlike $\mathrm{YBa_2Cu_3O_y}$ (YBCO) hole-doping
by oxygen loading in CLBLCO spans the complete $T_{c}(h)$ dome (for
$6.9<y<7.3$), ranging from the underdoped to the fully overdoped
regime.  $T_{c}^{max}$ is achieved at an optimum value of oxygen
content \cite{Keren2006}, $y^{opt}\sim7.13$.
Interestingly, $T_{c}^{max}$ can be tuned by varying the calcium
content, ranging between $0<x<0.4$ to shift $T_{c}$ from 45 K to 80 K.

Here we use the X-ray absorption spectroscopy (XAS), combined with
high resolution X-ray diffraction (HRXRD) to address the origin of the
variation of $T_{c}^{max}$ within the model system \CLBLCO.  Firstly
we deal with the question whether the variation of $T_{c}^{max}$ is
due to the change of the hole doping in the CuO$_2$ planes
\cite{Goldschmidt1993,Goldschmidt1995a,%Goldschmidt1995,Knizhnik1998,
Chmaissem2001}, followed by a study of the CuO$_2$ in-plane structural
modifications. We have performed Cu L (Cu $2p$) and O K-edge (O $1s$) XAS
to investigate the electronic structure near the Fermi level, probing
unoccupied Cu 3d and O 2p states (i.e., the hole states), while HRXRD
determines any variation in the lattice dimension and order.  The
results show that the large variation of $T_{c}^{max}$ with $x$ is not
due to the change of the hole density in the CuO$_2$ planes.  On the
other hand, there is a systematic variation of the lattice dimension
with $x$, implying that the increase of $T_{c}^{max}$ should be merely
due to changes in the chemical pressure on the CuO$_{2}$ sublattice.

%Figure 1
\begin{figure}
\includegraphics[width=0.40\textwidth]{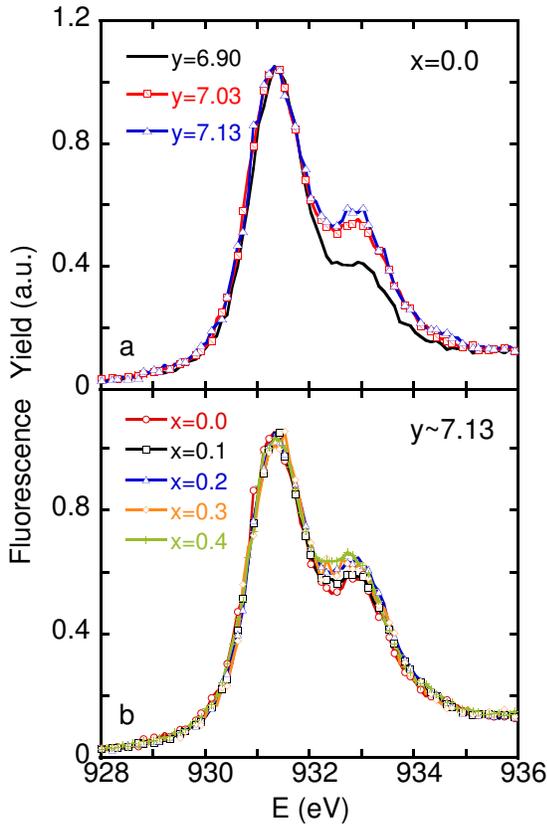}
\caption{\label{fig:CuLedge} (Color online) Comparison of the Cu L
edge XAS spectra of \CLBLCO, measured at T=20 K, for various O
contents, $y$, (upper panel) and Ca concentrations, $x$ (lower panel).
The introduction of oxygen produces an increase of the intensity of the
feature at 933 eV (ligand holes).  Negligible variations are observed
in the spectra as a function of Ca concentration.}
\end{figure}

%Figure 2
\begin{figure}
\includegraphics[width=0.40\textwidth]{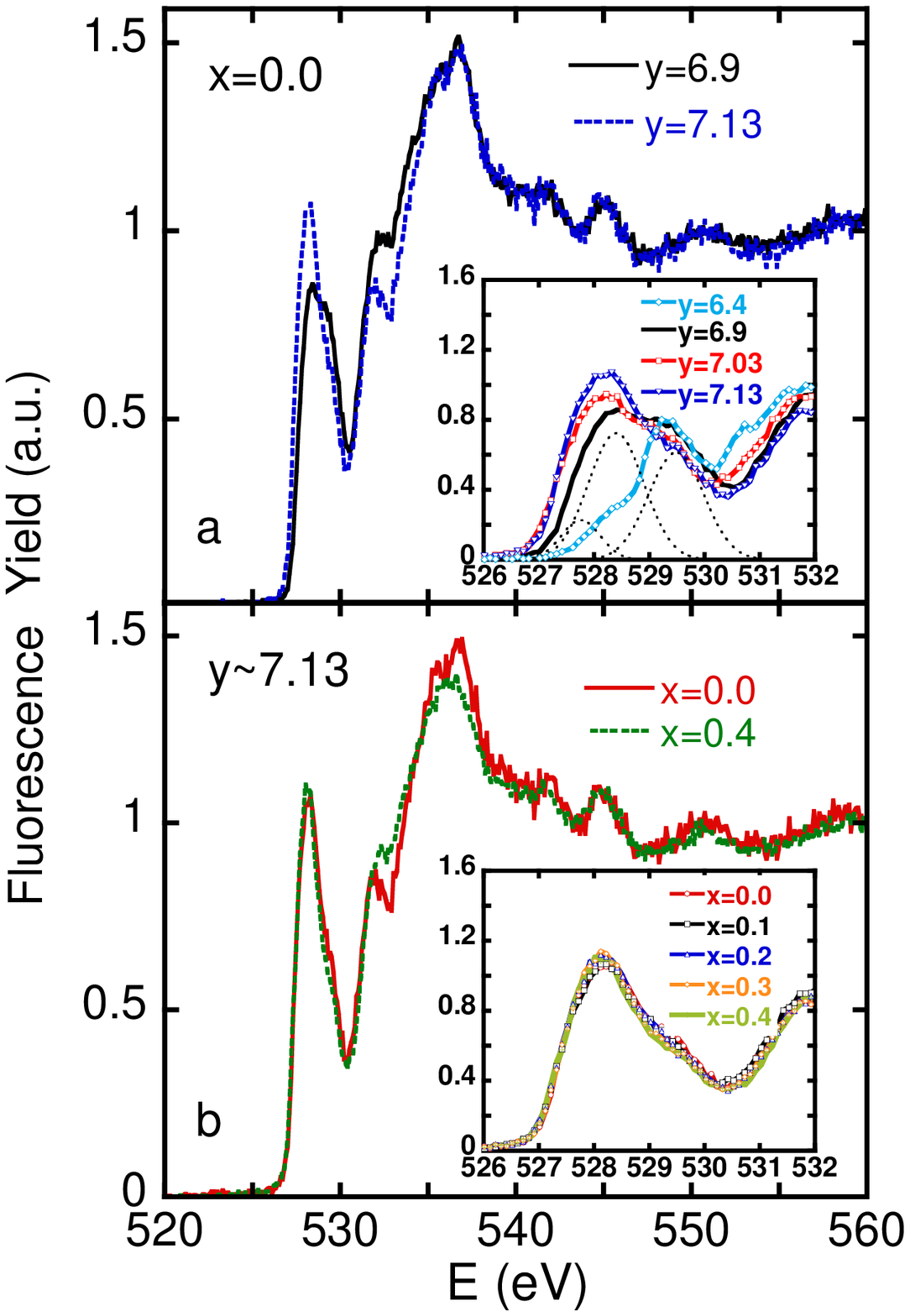}
\caption{\label{fig:OKedge} (Color online) Comparison of the O K edge
XAS spectra of \CLBLCO, measured at T=20 K, for various O contents,
$y$, (upper panel) and Ca concentrations, $x$, (lower panel).  These
spectra are normalised with respect to the atomic absorption about
70-80 eV above the absorption jump. Strong variations are evident in
the spectra as the oxygen content is changed, while a change of the Ca
content does not produce significant modifications. The insets are the
zooms over the pre-edge regions (see text).}
\end{figure}

The samples used for this work were prepared by the standard
solid-state synthesis method \cite{Goldschmidt1993}.  The oxygen was
varied by the topotactic technique \cite{Manca2001} to drive the
samples in the underdoped region and by annealing at 500$^o$C under an
oxygen pressure of $30-50$ atm to achieve the optimum $T_{c}^{max}$.
All the samples were characterized for their
phase purity and superconducting properties prior to the XAS and HRXRD
measurements.
The X-ray absorption spectra were collected at the ISIS beamline of
the BESSY storage ring (Berlin, Germany), with energy resolution of
$\sim$ 220 meV and $\sim$520 meV, respectively at 530 eV and at 930
eV. The measurements were performed at T=20 K by collecting
simultaneously the bulk sensitive fluorescence yield signal using a
windowless high purity Ge-detector, and the total electron yield by a
channeltron detector.  The base pressure in the analysis chamber was
better than 5$\cdot$10$^{-10}$ mbar during the measurements.  The
samples were scraped in-situ by a diamond file to get clean surfaces.
High resolution X-ray diffraction measurements were performed at the
beamline ID31 of the ESRF in Grenoble using a wavelength of
$\lambda$=0.4 {\AA}.

Fig. 1 displays the Cu L$_{3}$-XAS as a function of oxygen (upper)
and calcium (bottom) contents. The spectral shape is
similar to the one measured on the YBCO system
\cite{Nucker1995,Merz1998}.  Besides a white line at $\sim$931.3 eV
representing Cu $3d^9\rightarrow$ Cu $\underline{2p}\ 3d^{10}$
transition, a shoulder is observed at $\sim$932.8 eV due to Cu $3d^9
\underline{L} \rightarrow $ Cu $\underline{2p}~3d^{10} \underline{L}$
transition.  Here $\underline{L}$ indicates a hole in the oxygen
ligand orbital (O $2p$) induced by doping and the intensity of this
peak varies proportionally to the density of doping holes ($n_h$) per
Cu site \cite{Nucker1995,Merz1998,Salluzzo2008,Bianconi1987}.
A clear change in
$n_h$ due to the introduction of oxygen in the system is evident in
Fig. 1a. Conversely, Fig. 1b shows that Ca concentration (at fixed O
content) induces hardly any variation in the ligand hole peak
intensity. Unfortunately, since the Cu L$_{3}$-XAS spectra are
measured on polycrystalline samples, it is hard to distinguish the
contributions from chains and planes \cite{Nucker1995}. Nevertheless,
on the basis of the spectra shown in the Fig. 1b, it could be fairly
stated that the CLBLCO is globally isoelectronic, consistent with bond
valence sum calculations\cite{Chmaissem2001}.

In order to {\em selectively} detect the hole doping in the CuO$_2$
planes we measured the O K-XAS of CLBLCO (Fig.2). The pre-edge region
region for all the samples could be easily deconvoluted in three
Gaussians of similar width, centered at 527.6 eV, 528.4 eV and
529.5 eV. In the inset of the upper panel of Fig.2 the dashed
curves represent these three peaks for the sample with $x=0$
and $y=6.9$. Following the well established
peak assignment used for the XAS spectrum of YBCO, these peaks are
recognized respectively as : the "chain" plus apical (CA) oxygen, the
so-called Zhang-Rice (ZR) singlet and the upper Hubbard band (UHB)
contributions \cite{Nucker1995, Merz1998}. As expected, the spectral
weights of CA and ZR increase with oxygen, with a corresponding
reduction of the UHB signal.  In particular the variations of the ZR
peak are {\em directly} related to the change of hole density in the
CuO$_{2}$ plane \cite{Nucker1995,Merz1998}. On the other hand, if the
O content is fixed close to the optimum, with changing Ca
the pre-edge region of O K-XAS hardly shows any change (lower
panel of the Fig.2). This may be better seen in the inset, which
displays the $x$ dependence of the pre-edge region for the CLBLCO
samples close to $T_{c}^{max}$ ($y^{opt}\sim$7.13). It is noteworthy
that the shape of the O spectra at low energies, where the CA and the
ZR peaks appear, is unchanged by varying $x$. This is a {\em direct}
experimental evidence that no hole transfer occurs from the chain
sites to the planes. These results are in agreement with previous
bond-valence-sum calculations \cite{Chmaissem2001} and in contrast
to what was supposed in some previous works
\cite{Goldschmidt1993,Goldschmidt1995a}.

To further quantify the results, in  we have reported the
integrated difference $I-I_o$ (proportional to $n_h$) with respect
to the spectrum of a non superconducting composition near the
metal-insulator transition (i.e.,
$\mathrm{LaBa_{1.75}La_{0.25}Cu_3O_{6.9} }$) as a function of oxygen
(Fig.3a) and calcium (Fig.3b). The integration $I$ was carried
out in the energy region 520-528.5 eV and 920-936 eV for the O K and
Cu L$_{3}$-XAS respectively, and the results are consistent with each
other. It can be seen that the introduction of oxygen in the CLBLCO
system causes a rise of $n_h$ (Fig. 3a), as expected. On the other
hand, a change of Ca concentration does not produce a significant
variation (Fig. 3b). Furthermore, we can evaluate the inplane holes,
$n_{pl}$, by considering they are proportional to the intensity of the
ZR peak \cite{Nucker1995,Merz1998}, $n_{pl}\propto I_{ZR}$, and by
the re-scaling under the assumption that for the sample close to the
metal insulator transitions (y=6.9) is $n_{pl}\approx0.05$, and at
optimum oxygen (y$^{opt}\sim$7.13) is $n_{pl}\approx0.16$
\cite{Tallon1995}.  The data, displayed in the inset of fig.3b
as a function of the Ca content, show that at optimum oxygen $n_{pl}$
is constant within the uncertainties (maximum variation is
$\sim$0.02).  If we assume that the variation of $T_{c}^{max}$ from 45
K to 80 K is due to hole doping, for instance by using the parabolic
Tallon formula\cite{Tallon1995}, we should expect a
change of about 0.07 holes per CuO$_{2}$ plane, inconsistent with the
present experimental result.

%Figure 3
\begin{figure}
\includegraphics[width=0.7\columnwidth]{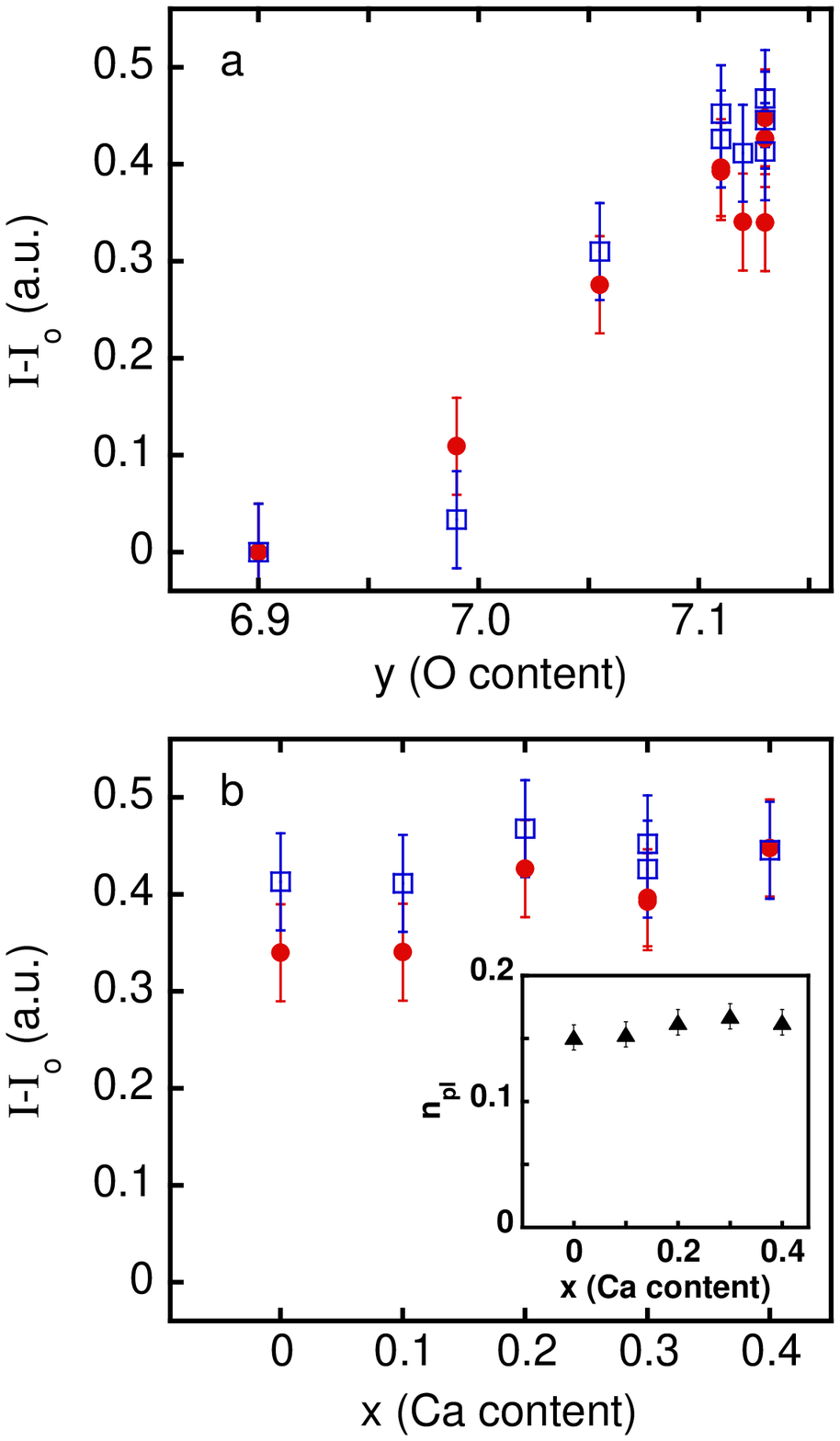}
\caption{\label{fig:holes} (Color online) O content, $y$, (upper
panel) and Ca content, $x$, (lower panel) dependence of the integrated
XAS spectral difference (proportional to the hole density per Cu site,
$n_h$), measured at T=20 K. The Cu L-edge (filled circles) and O
K-edge (empty squares) XAS spectra of \CLBLCO are consistent with each
other showing that the hole number increases as oxygen is introduced,
while it remains nearly constant as the Ca concentration is
varied. The inset (panel b) shows the inplane holes, $n_{pl}$,
proportional to the intensity of the ZR peak, appearing almost
constant with $x$, i.e. no hole transfer takes place from ``chain'' to
planes while $T_{c}^{max}$ varies from 45 K to 80 K.}
\end{figure}

To summarize, the results construct a clear evidence that not only
the total number of holes is constant, but also the holes in the
CuO2 planes do not change appreciably with x at $y^{opt}\sim$7.13
in CLBLCO system. Thus it is reasonable to conclude that a large
change of $T_{c}^{max}$ from 45 K to 80 K as a function of $x$ cannot
be due to the variation of the in-plane hole density. Therefore, apart
from hole density, there should be other parameter which controls the
$T_{c}$ in cuprates.  Other responsible parameters could be either
substitutional order/disorder and/or the chemical pressure. As far as
the former is concerned, it has been revealed that the substitutional
disorder does not change or slightly increases \cite{Keren2006} with
$x$. Therefore, T$_{c}$ should be either constant or even reduced
\cite{Fujita2005}, inconsistent with the observed behaviour. Thus the
increase of T$_{c}$ with $x$ should be somehow related with the
chemical pressure.

%Figure 4
\begin{figure}
\includegraphics[width=0.8\columnwidth]{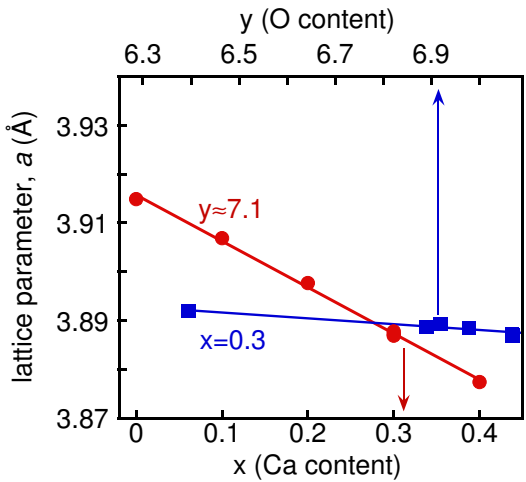}
\caption{\label{fig:xr} (Color online) In plane lattice parameter $a$
of \CLBLCO\ at $y\sim7.13$ vs. calcium content (bottom axis), and at
$x=0.3$ vs. oxygen content (top axis). The lattice parameter $a$,
which is directly related with the chemical pressure on the CuO$_2$
layers, changes as the Ca concentration is modified, while remains
nearly constant as oxygen is introduced.}
\end{figure}

%Figure 5
\begin{figure}
\includegraphics[width=0.8\columnwidth]{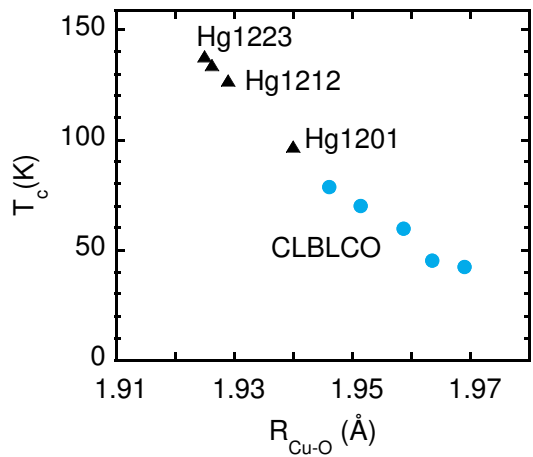}
\caption{\label{fig:RCuO} (Color online)
$T_{c}^{max}$ as a function of the Cu-O inplane distance,
$R_{Cu-O}=a/(2cos\theta)$ (where $\theta$ is the buckling angle of the
CuO$_2$ planes). The data for the Hg-cuprate family are from
Ref.\cite{Lokshin2001}.}
\end{figure}

Here we should recall that there is a lattice mismatch between the CuO$_2$
layer and the rock-salt block with a large compressive stress on
the CuO$_{2}$ plane. This lattice mismatch has been proposed
\cite{Bianconi2000, Agrestini2003,Bianconi2001,Fratini2008} as the third variable needed to
define the phase diagram of cuprate perovskites, beyond
superconducting temperature $T_{c}$ and hole doping.  In CLBLCO the
insertion of calcium cations is accompanied by the substitution of
lanthanum at the barium sites which produces an internal mismatch
pressure, due to the large difference in ionic radii (R$_{Ca(II)}$=112
pm, R$_{La(III)}$=116 pm, and R$_{Ba(II)}$ =142 pm
\cite{Shannon1976}). The variation of this chemical pressure
may be directly determined by measuring the structural parameters
(lattice spacing, buckling of the planes, etc.) as a function of $x$.
We have performed a high resolution powder X-ray diffraction (HRXRD) study to
measure the effect of Ca doping on the lattice structure. Rietveld
analysis did not reveal any structural transition and the atomic
parameters are in good agreement with those in the literature
\cite{Chmaissem1999,Chmaissem2001}. In particular, we measured the
in-plane lattice parameter, $a$, which is directly linked to the Cu-O
distance of the CuO$_2$ plaquette. Fig.4 shows the evolution of the
$a$-axis for the samples with the $T_{c}^{max}$ ($y\sim$7.13) as a
function of $x$ (bottom axis), compared with the samples at $x=0.3$ as
a function of the oxygen content $y$ (top axis). The chemical mismatch
pressure remains almost constant when the O content is changed, while
it shows a substantial variation as a function of $x$, i.e., when the
$\mathrm{Ca\rightarrow La\rightarrow Ba}$ substitutions occur.  The
main effect of the inplane chemical mismatch is the variation of the
Cu-O distance, $R_{Cu-O}=a/(2cos\theta)$. Here $\theta$ is the buckling
angle of the CuO$_2$ planes, evaluated by the HRXR diffraction data, and ranges from 5.0(3) to 6.2(3) degrees going from x=0.4 to x=0, in agreement with ref. \cite{Chmaissem1999}. Note that this corresponds to a negligible correction to the Cu-O distance (of the order of 0.003 Amstrong).
It is well known that the buckling affects the superconducting properties of cuprates \cite {Buchner1994,Dabrowski1996}. However, in CLBLCO the relation between T$_c$ and the buckling angle is complex; with oxygen it increases, while an opposite behavior is observed when the calcium content is varied \cite{Chmaissem1999}. In any case it should be considered that the change of T$_c$ as a function of $\theta$ for the La$_2$CuO$_4$ family (maximum T$_c$=45K) is found to be about 3K/Deg (e.g., see fig.3 of Ref. \cite{Dabrowski1996}). In the CLBLCO, the change is about 1 Degree from x=0.4 to x=0 that, even renormalizing for the maximum T$_c$=80K, should lead to a little T$_c$ reduction of about 6K. This is about 6 times smaller than what has been observed, i.e. a change from 80K to 45K. Thus, the variation of the buckling angle cannot explain the T$_c$ changes in the present system as a function of $x$.

Fig.5 shows $T_{c}^{max}$ of CLBLCO as a function of
$R_{Cu-O}$ following the same trend of the Hg-cuprate family \cite{Lokshin2001}
(note that also in this last family the buckling is irrelevant since
the CuO$_2$ planes are flat). The regular scaling of T$_c$ with the Cu-O
distance provides clear evidence that the chemical pressure is a key parameter
for the superconducting ground state of cuprates. This has been also seen in the new Fe-based superconductors \cite{Ren2008}.
We do not know what is the intimate mechanism through which the chemical
pressure affects the superconductivity, but it can be hypothesized that the
chemical pressure tunes somehow the coupling strength among the Cooper pairs.

In summary, we have studied the problem of variation of $T_{c}^{max}$
using X-ray absorption spectroscopy and high resolution X-ray
diffraction on the charge compensated \CLBLCO\ (0$\leq x<$0.5,
$y\approx$7.13), a peculiar system that permits to explore
independently the effect of two parameters, i.e. the hole density and
the chemical pressure. The O K and Cu L$_{3}$ XAS provide consistent
results with negligible spectral change upon $x$, suggesting that the
large change of $T_{c}^{max}$ with cationic substitution is not
related with the variation of the hole density in the CuO$_2$ planes.
The high resolution X-ray diffraction data reveal a systematic
variation in the lattice parameter due to varying the chemical
pressure on the CuO$_{2}$ sublattice due to cationic substitution in
the rock-salt layers of the title system. Since there is a large
increase of $T_{c}^{max}$ with the cationic substitution without any
appreciable change in the hole density, the only possibility is that
the chemical pressure on the CuO$_{2}$ sublattice being the
controlling parameter which could act cooperatively with the hole
density.  Therefore, the results have direct implication on the
superconducting ground state of the cuprate perovskites, suggesting
that, besides the hole density, the chemical pressure is the second
variable that controls the superconducting critical temperature.
Although the intimate mechanism which links the chemical pressure to
the superconductivity is still unclear, these results provide an
important experimental feedback. To further enlighten this problem,
CLBLCO single crystals are highly desirable since this compound
permits to control $T_c$ at constant doping and within the same
cuprate family.

{\bf Acknowledgements} We gratefully acknowledge ESRF and BESSY for
provision of beamtime and G.Mitdank and A.Ariffin for the
assistance. We thank G.Concas and E.Pusceddu for preliminary sample
preparation and M.Filippi for help in HRXRD diffraction measurements.
This work was supported by PRIN06 ``Search for critical parameters in
high $T_c$ cuprates'' and by the European Community-Research
Infrastructure Action under the FP6 ``Structuring the European
Research Area" program (IA-SFS, Contract No. RII3-CT-2004-506008).
The authors also acknowledge a partial support from the European
project 517039 ``Controlling Mesoscopic Phase Separation" (COMEPHS)
(2005). We are grateful to A. Bianconi and A. Keren for fruitful
discussions.

\bibliography{CL}

\end{document}